\documentclass{article}
\usepackage{spconf,amsmath,graphicx}
\usepackage{amssymb}
\usepackage{booktabs}
\usepackage{amsmath,bm}
\usepackage{algorithm}
\usepackage{algorithmic}

\usepackage{soul}
\usepackage{multirow}
\usepackage{colortbl}

\usepackage[hyphens]{url} 
\usepackage{cite}
\usepackage{hyperref}
\title{WiFi-based Spatiotemporal Human Action Perception}
\name{
    Yanling Hao$^{\star}$ \qquad Zhiyuan Shi$^{\dagger}$ \qquad
    Yuanwei Liu$^{\star}$
}
 \address{
    $^{\star}$ Queen Mary University of London, London, UK\\
     $^{\dagger}$ Onfido Research London, London, UK\\
    \{yanling.hao,yuanwei.liu\}@qmul.ac.uk, zhiyuan.shi@onfido.com
 }

\begin{document}
%
\maketitle
\begin{abstract}
 WiFi-based sensing for human activity recognition (HAR) has recently become a hot topic as it brings great benefits when compared with video-based HAR, such as eliminating the demands of line-of-sight (LOS) and preserving privacy. Making the WiFi signals to 'see' the action, however, is quite coarse and thus still in its infancy. An end-to-end spatiotemporal WiFi signal neural network (STWNN) is proposed to enable WiFi-only sensing in both line-of-sight and non-line-of-sight scenarios. Especially, the 3D convolution module is able to explore the spatiotemporal continuity of WiFi signals, and the feature self-attention module can explicitly maintain dominant features. In addition, a novel 3D representation for WiFi signals is designed to preserve multi-scale spatiotemporal information. Furthermore, a small wireless-vision dataset (WVAR) is synchronously collected to extend the potential of STWNN to 'see' through occlusions. Quantitative and qualitative results on WVAR and the other three public benchmark datasets demonstrate the effectiveness of our approach on both accuracy and shift consistency.
\end{abstract}
\section{Introduction}
\label{sec:intro}
Recent years have witnessed increasing research interest in human action recognition (HAR) as it expands sensing areas and provides vast potential applications~\cite{isogawa2020optical,luo2021intelligent} in various sensing scenarios, such as assisted living~\cite{wu2018wifi}, health monitoring~\cite{tan2018exploiting}, surveillance~\cite{wang2018csi}, etc. Many new action sensing technologies~\cite{isogawa2020optical,luo2021intelligent} are continuously emerging, which enlarges signal acquisition range and enriches measurement data~\cite{luo2021intelligent}. These sensing techniques motivate the breakthrough of long-time monitoring in the non-intrusive pattern~\cite{AAAAli2019making}. 

Video-based systems demand the coverage area within line-of-sight(LOS)~\cite{isogawa2020optical}. In general, lighting variation affects the quality of images and thus, analyzed information~\cite{beddiar2020vision}. In the same way, perspective change, especially using a single view acquisition device, provides a limited visualization of activities being analyzed. This induces occlusion with its different types: self-occlusion where body parts occlude each other, occlusion of another object, and partial occlusion~\cite{beddiar2020vision}. From a user perspective, the presence of cameras can affect privacy and cannot be employed in many environments. Therefore, a passive monitoring system based on RF sensing is a more sound way to sidestep such drawbacks~\cite{luo2021intelligent}. 
\begin{figure}[t]
    \vspace{-5pt}
    \centering
    \includegraphics[scale=0.26]{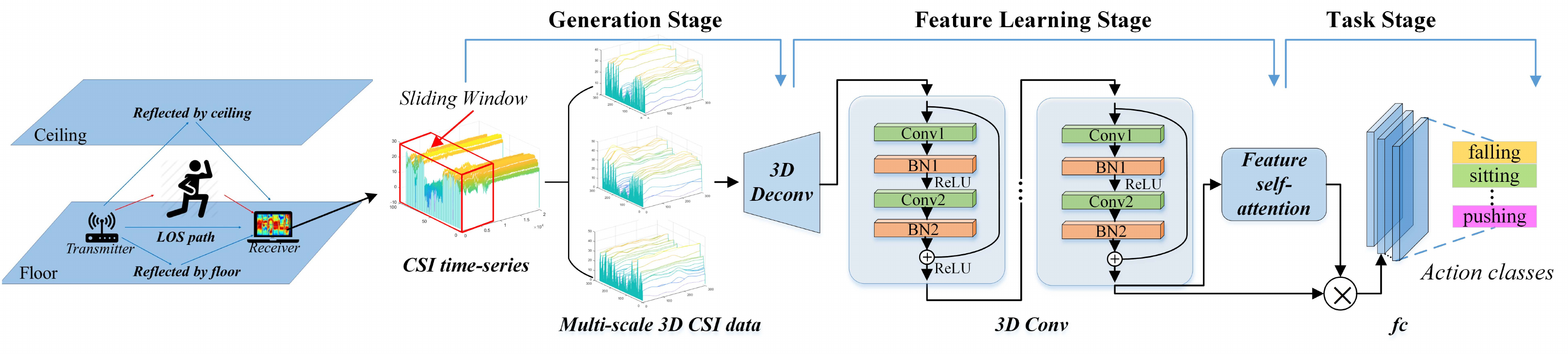}
    \caption{The flowchart of proposed STWNN}
    \label{fig:architecture}
\vspace{-10pt}
\end{figure}
Unlike video-based solutions, RF especially WiFi channel state information (CSI) based sensing is insensitive to lighting conditions and not intrusive without the privacy issue~\cite{li2021deep}. WiFi-based solutions have no requirements of LOS thereby enabling larger detection areas. Existing systems, however, are quite coarse~\cite{fang2020person}. Past systems focus primarily on manually designed features, dependent on prior knowledge and incapable of adequately mining spatiotemporal information in CSI streams~\cite{alazrai2020dataset}. Furthermore, separate stages for feature extraction and classifier learning may reduce the accuracy of recognition results. Therefore, it is worth exploring how to non-manually obtain spatiotemporal features and jointly optimize feature learning as well as the classification process.

To address the problems, we propose an end-to-end spatiotemporal WiFi-based neural network (STWNN) to exploit the spatiotemporal characteristics of CSI signals simultaneously. To summarize, our contributions are 1) We propose a novel method for representing WiFi signals in a multi-scale 3D spatiotemporal form; 2) We design a 3D convolution module and attention module to exploit the inherent spatial, temporal, and frequency features and embedded in a residual manner to reduce training burden; 3) We collect synchronous video and WiFi datasets (WVAR) to enable STWNN to “see” through the occlusions; 4) We conduct experiments on three public benchmark datasets. The results show that our method outperforms competitive baselines with a good margin on the classification accuracy.

\section{The Proposed spatiotemporal Neural Network}
\label{sec:metho}



\subsection{Generation stage}
\subsubsection{Channel state information}
\label{csi_gen}
The WiFi-based sensing principle is leveraging the influence of perceptual targets on the transmitted signal for recognition~\cite{wang2016rt}. Generally, a WiFi system can be modeled as follows:
\begin{equation}
{B_s}(i) = {H_s}(i){A_s}(i) + \theta,
\end{equation}
where $s\in[1,\cdots,N_s]$ represents the index of the orthogonal frequency-division multiplexing (OFDM) subcarriers employed in the WiFi device, ${N_s}$ is the total number of the OFDM subcarriers, $i$ represents the index of the transmitted and received packets, ${\bm{A}_s}(i)$ and ${\bm{B}_s}(i)$ are the ${i^{th}}$ transmitted and received packets associated with the OFDM subcarrier frequency ${s}$, respectively, ${\bm{\theta}}$ represents the received noise, and ${\bm{H}_s}$ is a complex-valued matrix of dimensions ${N_T} \times {N_R}$ that comprises the CSI measurements for the OFDM subcarrier frequency $s$. ${N_T}$ and ${N_R}$ represent the number of transmitting and receiving antennas, respectively.

\subsubsection{Multi-scale 3D CSI data generation}
\label{multi_scale}
 \begin{figure}[h]
    \vspace{-5pt}
    \centering
    \includegraphics[scale=0.18]{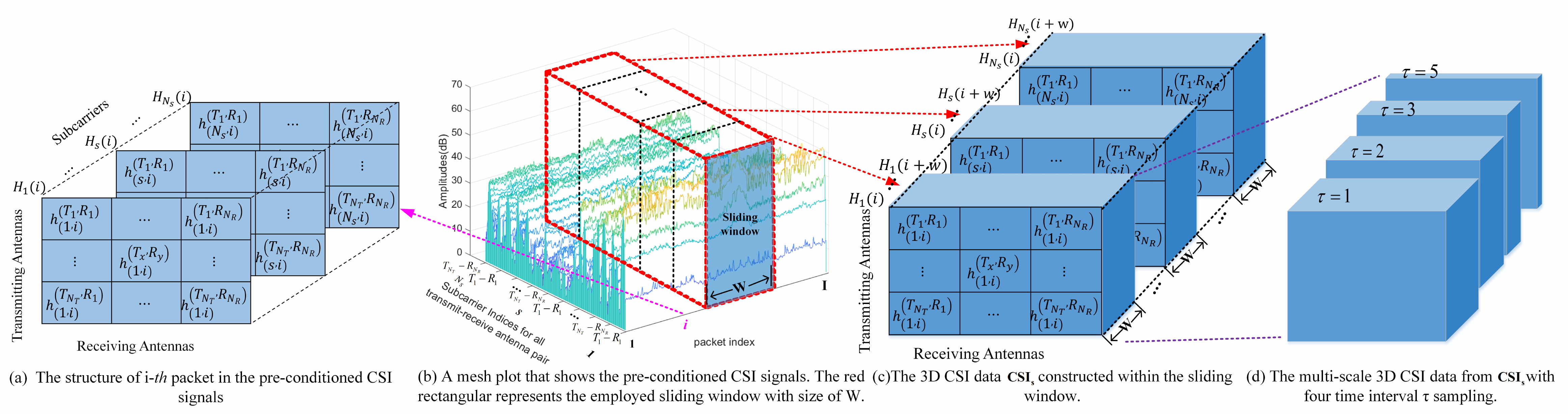}
    \caption{The construction procedure of a multi-scale 3D data.}
    \label{fig:csi_3D}
\vspace{-10pt}
\end{figure}
The objective of the generation stage is to convert the 4D pre-conditioned CSI signals into a set of 3D CSI data samples that preserve the time, spatial as well as frequency information comprised within the CSI signals. As seen in Fig.~\ref{fig:csi_3D}(b), the CSI signals included in each time serial are arranged in a manner of 4D matrix $M \times I$, where $M = {N_T} \times {N_R} \times {N_S}$, and I represents the number of packets recorded in a particular CSI time serial. A sliding window along the time axis is utilized to divide the CSI signals into a set of overlapped segments, as illustrated in Fig.~\ref{fig:csi_3D}(b). The size of each segment is set to $W$ packets and the overlap between every two consecutive segments is set to $\varpi $, where $\varpi  \le W$. The s-$th$ CSI segment $CSI_s(i)$ is $W \times {N_T} \times {N_R}$, where $s \in [1, \ldots ,{N_S}]$. Fig.~\ref{fig:csi_3D}(c) illustrates the construction procedure of constructing a 3D data $\boldsymbol{CSI}(i)$. To have a multi-scale view of the CSI signals for the complex actions, various time intervals $\tau $ are set to sample the CSI data. Finally, the 3D CSI samples are up-sampled to be processed by 3D convolution modules, and the details are listed in Section~\ref{sec:exper}.

\subsection{Feature learning stage}
\subsubsection{3D convolution module}
Convolutional neural networks with 3D kernels can directly extract spatiotemporal features from videos, however, suffer from the heavy training burden due to a large number of their parameters. To mitigate the issue, we construct the network based on ResNet, which introduces shortcut connections that bypass a signal from one layer to the next. The connections pass through the gradient flows of networks from later layers to early layers, and ease the training of very deep networks. The connections bypass a signal from the top of the block to the tail. Our 3D convolution module consists of multiple residual blocks seen in Fig.~\ref{fig:architecture}.

\subsubsection{Feature self-attention module}
Inspired by the attention mechanism, our works formulate attention drift as a sequential process to capture different attended aspects. The learned sequential features by the 3D convolution module will be employed as the inputs of the attention model as self-attention with no prior information, as seen in Fig.~\ref{fig:architecture}. Given n feature vectors $\boldsymbol{\alpha_i}{\rm{ }}$, $i = {\rm{ }}1,2,\ldots,n $ derived from the 3D convolution module, a score function $\Phi ( \cdot )$ such as tanh, relu and linear evaluates the importance of each feature vector by calculating a score ${\beta _i}$ as follows:
\begin{equation}
    {\beta _i} = \Phi ({{\boldsymbol{\chi}}^T}{\boldsymbol{\alpha }_{\bf{i}}} + b),
\end{equation}
where $\chi^T$ and $b$ are weight vector and bias respectively. After obtaining the score for each feature vector, we can normalize it using the softmax function. The final Mask of the attention model is the multiplication of the feature vectors and their normalized scores, which is shown as follows:
\begin{equation}
       Mask = \sum\limits_{i = 1}^n {(softmax({\beta _i}) * {\boldsymbol{\alpha _i}}} ) = \sum\limits_{i = 1}^n {(\frac{{\exp ({\beta _i})}}{{\sum\nolimits_i {\exp ({\beta _i})} }} * {\boldsymbol{\alpha _i}}} ).
\end{equation}

\vspace{-10pt}
\subsection{Task stage}
The task stage is to leverage multi-scale spatiotemporal features learned above to compute the outputs for a specific task. Cross-entropy loss is a basic option to measure the difference between the network outputs O and the ground-truth values as follows.
\begin{equation}
  { \mathcal{L}} =  - \lambda \sum\limits_{j = 1}^J {{G^j}\log (Mas{k^j}*{O^j})}  - (1 - \lambda )\sum\limits_{j = 1}^J {{G^j}\log ({O^j})} 
\end{equation}
where * is the convolution operation, $J$ is the snippet number of training samples, and $\lambda$ is the weight coefficient. A typical value is $\lambda = 0.5$  in our experiments. In addition, we utilize the Stochastic Gradient Descent with Momentum to train the parameters.
\begin{figure}[t]
    \setlength{\abovecaptionskip}{-25pt}
    \centering
    \includegraphics[scale=0.36]{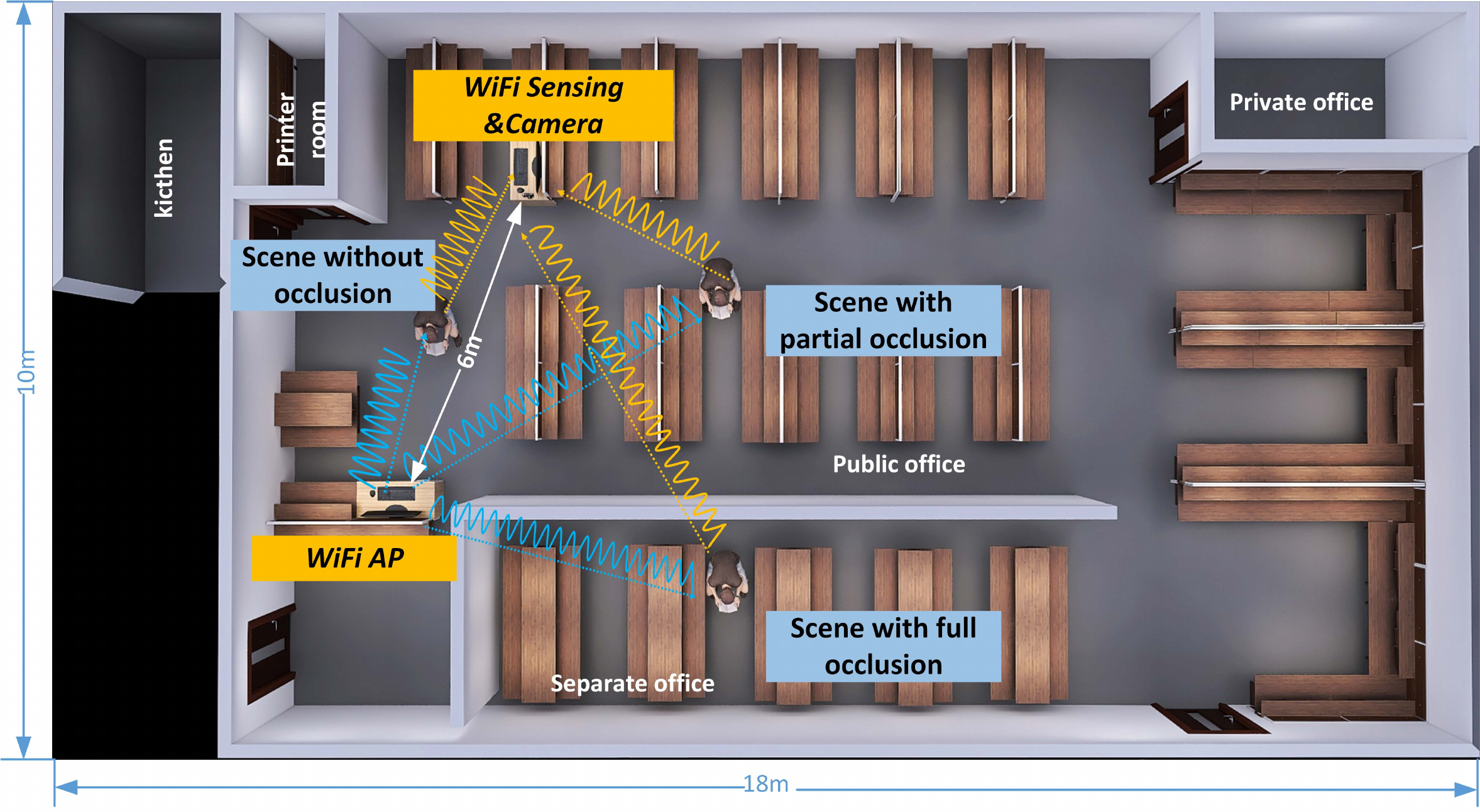}
    \caption{Three experiment scenes in indoor environments.}
    \label{fig:scene}
\vspace{-15pt}
\end{figure}
\section{Experiments}
\label{sec:exper}
\noindent\textbf{Our WVAR dataset}. WVAR collection was conducted in one spacious office apartment as shown in Fig. \ref{fig:scene}. Two volunteers were asked to implement nice activities and repeat for five trials in different motion details such as varied directions to ensure the diversity of the actions. The experimental hardware consists of two desktop computers as transmitter and receiver operating in IEEE 802.11n monitor mode at 5.4 GHz with a sampling rate of 100 Hz. The subcarriers ${N_S}$ are equal to 30 and 3 antennas both in transmitter ${N_T}$ and receiver ${N_R}$ are activated. We employed the CSI extraction tool\footnote {It is available at \url{https://github.com/dhalperi/linux-80211n-csitool-supplementary}} for CSI signals recording and CSI packets extraction. To synchronize the images and wireless data, a deep camera D435i\footnote{It is available at \url{https://www.intelrealsense.com/depth-camera-d435i/}} was attached to our receiver desktop at the same location as the wireless card. We recorded the video at 20 FPS, i.e. every five CSI samples corresponding to one frame in the video.
\begin{figure}[h]
    \centering
    \includegraphics[scale=0.3]{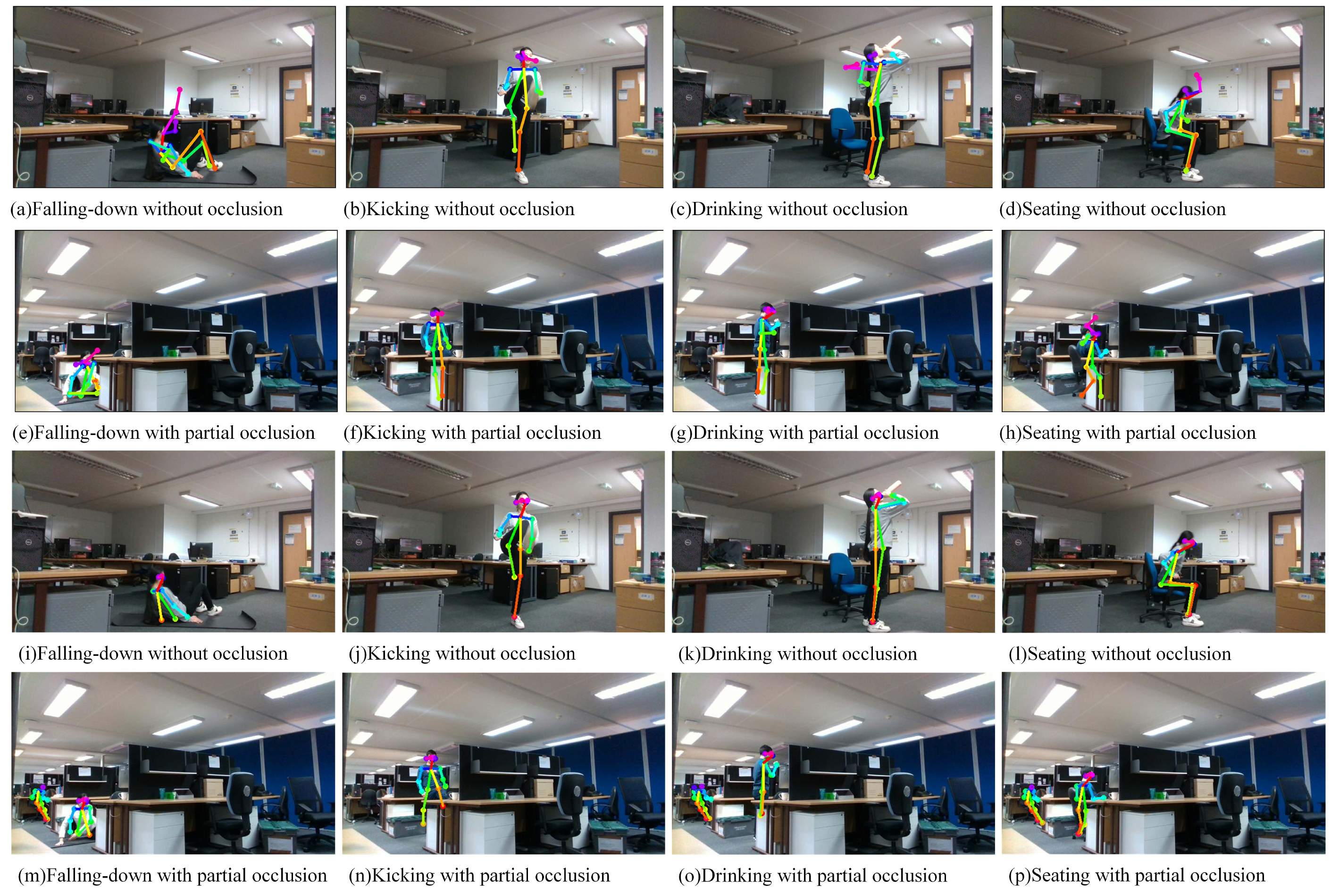}
    \caption{(a)-(h) WiFi-based and (i)-(p) video-based skeleton results on the dataset WVAR in scenes without occlusion, and with partial occlusion. In the scene without occlusion as the first two rows show, the skeleton results by WiFi are comparable in seating, and better in a self-occlusion case like falling-down than those by video. As for the scene without occlusion in the last two rows, the skeleton results by WiFi are more precise seen in the legs in (g) compared to (o), and have less false detection like the chairs than those by video.}
    \label{fig:skeleton}
\vspace{-10pt}
\end{figure}

\vspace{-5pt}
\noindent\textbf{HHI, CSLOS and WAR dataset}. The dataset HHIs~\cite{alazrai2020dataset} comprises 12 different interactions performed by 40 distinct pairs of subjects while performing different human-to-human interactions (HHI) inside an office with 10 different trials. Another cross-scene dataset (CSLOS)~\cite{baha2020dataset} is provided by the same group as the HHI. CSLOS is comprised of five experiments performed by 30 different subjects in two LOS environments. Each subject performed 20 trials for each of the experiments with different variations of human movements. The dataset WAR~\cite{yousefi2017survey} consists of 6 persons for 6 activities with 20 trials for each in an indoor office. The sampling rate is 1 kHz. 

\noindent\textbf{Baselines}. We design a 2D baseline with the same neural network structure as STWNN (2DWNN). Besides, the classic SVM~\cite{crammer2001algorithmic} is deployed for comparison. 

\noindent\textbf{Quantitative Results}. Table~\ref{tab:WVAR} shows the classification accuracy of the WVAR dataset. We tested the data from all scenarios (All), the scenes with partial (S-p) and full occlusion (S-f), respectively. For all scenarios, SVM (non$\_$gen) without the generation stage performs worse than SVM, indicating the generation stage's effectiveness. The overall accuracy (OA) of STWNN is higher than these of others and over 85$\%$ on all the actions. It is possible that all activities have obvious trajectories in the spatial domain over time. STWNN can pay attention to the characteristics of both spatiotemporal domains. For the scene with partial and full occlusions, STWNN surpasses the other two methods with a margin of around 6$\%$.These results indicate that robust to the influence of the environmental disturbance.

\begin{table}
\caption{Classification accuracy on the dataset WVAR.}
\label{tab:WVAR}
\tiny
\scalebox{0.8}{
\centering
\begin{tabular}{l!{\vrule width \lightrulewidth}l!{\vrule width \lightrulewidth}lcccccccccc} 
\toprule
\textbf{Scene}                                                                                          & \textbf{Method}       & \textbf{\begin{tabular}[c]{@{}l@{}}\textbf{fall}\\ \textbf{\_down}\end{tabular}} & \textbf{throw} & \textbf{push} & \textbf{kick} & \textbf{punch} & \textbf{jump} & \textbf{\begin{tabular}[c]{@{}l@{}}\textbf{phone}\\ \textbf{\_talk}\end{tabular}} & \textbf{seat} & \textbf{drink} & \textbf{OA}    \\ 
\midrule
\multirow{5}{*}{\textbf{All}}                                                                     & \textbf{\begin{tabular}[c]{@{}l@{}}\textbf{SVM}\\ \textbf{(non\_gen)}\end{tabular}} & 0.95               & 0.80           & 0.81          & 0.80          & 0.47           & 0.71          & 0.73                 & \textbf{0.93} & 0.94           & 0.79           \\
                                                                                                         & \textbf{SVM}           & \textbf{1.00}      & 0.92           & 0.90          & 0.94          & 0.93           & 0.94          & 0.91                 & 0.88          & \textbf{1.00}  & 0.94           \\
                                                                                                         & \textbf{2DWNN}         & \textbf{1.00}      & \textbf{1.00}  & \textbf{1.00} & 0.86          & 0.88           & \textbf{1.00} & \textbf{1.00}        & 0.81          & \textbf{1.00}  & 0.94           \\
                                                                                                         & \textbf{\begin{tabular}[c]{@{}l@{}}\textbf{STWNN}\\\textbf{(skeleton)}\end{tabular}
                                                                                                         }           & \textbf{1.00}      & 0.88           & \textbf{1.00}          & 0.85          & 0.90           & \textbf{1.00}          & \textbf{1.00}          & 0.89          & \textbf{1.00}  & 0.94           \\
                                                                                                         & \textbf{STWNN}         & 0.96               & 0.96           & \textbf{1.00} & \textbf{0.97} & \textbf{1.00}  & 0.97          & 0.91                 & 0.85          & 0.97           & \textbf{0.95}  \\
                                                                                                         &                        &                    &                &               &               &                &               &                      &               &                &                \\
\multirow{3}{*}{\begin{tabular}[c]{@{}l@{}}\textbf{S-p}\end{tabular}} & \textbf{SVM}           & 0.97      & 0.98  & \textbf{0.99} & 0.99 & 0.75           & 0.71          & 0.75                 & \textbf{1.00} & \textbf{1.00}  & 0.90           \\
                                                                                                         & \textbf{2DWNN}         & 0.83               & 0.98  & 0.97 & \textbf{1.00} & 0.95  & 0.75          & \textbf{0.99}        & 0.83          & \textbf{1.00}  & 0.92           \\
                                                                                                         & \textbf{STWNN}         & \textbf{1.00}      & \textbf{0.99}  & 0.94 & 0.98 & \textbf{0.96}  & \textbf{0.97} & \textbf{0.99}        & 0.98 & 0.85           & \textbf{0.96}  \\
                                                                                                         &                        &                    &                &               &               &                &               &                      &               &                &                \\
\multirow{3}{*}{\begin{tabular}[c]{@{}l@{}}\textbf{S-f}\end{tabular}}    & \textbf{SVM}           & \textbf{1.00}      & \textbf{1.00}  & 0.75          & \textbf{1.00} & \textbf{1.00}  & 0.55          & 0.75                 & 0.99 & 0.99  & 0.85           \\
                                                                                                         & \textbf{2DWNN}         & 0.97      & 0.80           & \textbf{1.00} & 0.86          & 0.98  & \textbf{0.86} & \textbf{1.00}        & \textbf{1.00} & 0.71           & 0.89           \\
                                                                                                         & \textbf{STWNN}         & \textbf{1.00}      & 0.86           & \textbf{1.00} & 0.98 & \textbf{1.00}  & 0.75          & \textbf{1.00}        & 0.98 & \textbf{1.00}  & \textbf{0.94}  \\
\bottomrule
\end{tabular}}
\vspace{-15pt}
\end{table}
\begin{table}[h]
\vspace{-10pt}
\caption{Classification accuracy on the dataset WAR.}
\label{tab:WAR}
\centering
\scriptsize
\begin{tabular}{@{}lccccccc@{}}
\toprule
\textbf{Methods}     & \textbf{lie\_down} & \textbf{fall} & \textbf{run}  & \textbf{sit\_down} & \textbf{stand\_up} & \textbf{walk} & \textbf{OA}   \\ \midrule
\textbf{RF}~\cite{ho1995random}   & 0.53              & 0.60          & 0.81          & 0.88              & 0.49              & 0.57          & 0.65          \\
\textbf{HMM}~\cite{eddy2004hidden}  & 0.52              & 0.72          & 0.92          & 0.96              & 0.76              & 0.52          & 0.73          \\
\textbf{LSTM}~\cite{yousefi2017survey} & 0.95              & 0.94          & \textbf{0.97} & 0.81              & 0.83              & \textbf{0.93} & 0.91          \\
\textbf{SVM}         & 0.91              & 0.96          & 0.93          & 0.96              & 0.71              & 0.87          & 0.93          \\
\textbf{2DWNN}      & 0.93              & 0.93          & 0.93          & \textbf{0.98}     & 0.90              & 0.86          & 0.95          \\
\textbf{STWNN}      & \textbf{0.96}     & \textbf{0.99} & \textbf{0.97} & 0.97              & \textbf{0.96}     & \textbf{0.93} & \textbf{0.97} \\ \bottomrule
\end{tabular}
\vspace{-10pt}
\end{table}

\begin{table*}[h]
\caption{Classification accuracy on the dataset HHI.}
\label{tab:HHI}
\resizebox{\textwidth}{!}{
\begin{tabular}{@{}lccccccccccccc@{}}
\toprule
\textbf{Methods}           & \textbf{approaching} & \textbf{departing} & \textbf{\begin{tabular}[c]{@{}c@{}}hand\_\\ shaking\end{tabular}} & \textbf{high five} & \textbf{hugging} & \textbf{\begin{tabular}[c]{@{}c@{}}kicking\_\\ left\_leg\end{tabular}} & \textbf{\begin{tabular}[c]{@{}c@{}}kicking \_\\ right\_leg\end{tabular}} & \textbf{\begin{tabular}[c]{@{}c@{}}pointing\_\\ left\_hand\end{tabular}} & \textbf{\begin{tabular}[c]{@{}c@{}}pointing\_ \\ right\_hand\end{tabular}} & \textbf{\begin{tabular}[c]{@{}c@{}}punching\_ \\ left\_hand\end{tabular}} & \textbf{\begin{tabular}[c]{@{}c@{}}punching\_\\  right\_hand\end{tabular}} & \textbf{pushing} & \textbf{OA}   \\ \midrule
\textbf{GoogleNet}~\cite{szegedy2015going}   & 0.93                 & 0.93               & 0.79                 & 0.76               & 0.64             & 0.54                                                                   & 0.50                                                                     & 0.78                                                                     & 0.77                                                                       & 0.59                                                                      & 0.59                                                                       & 0.68             & 0.71          \\
\textbf{ResNet-18}~\cite{he2016deep}   & 0.92                 & 0.90               & 0.85                 & 0.79               & 0.77             & 0.68                                                                   & 0.60                                                                     & 0.82                                                                     & 0.80                                                                       & 0.60                                                                      & 0.65                                                                       & 0.76             & 0.76          \\
\textbf{Squeeze-Net}~\cite{iandola2016squeezenet}   & 0.95                 & 0.93               & 0.83                 & 0.76               & 0.70             & 0.66                                                                   & 0.62                                                                     & 0.78                                                                     & 0.79                                                                       & 0.60                                                                      & 0.72                                                                       & 0.74             & 0.76          \\
\textbf{E2EDLF}~\cite{alazrai2020end}    & 0.96                 & 0.92               & 0.89                 & 0.84               & \textbf{0.86}    & 0.78                                                          & 0.82                                                            & 0.85                                                            & 0.90                                                              & 0.73                                                                      & \textbf{0.80}                                                              & 0.86    & 0.85 \\
\textbf{SVM}               & \textbf{0.99}        & 0.96               & 0.90        & 0.83               & 0.82             & 0.73                                                                   & 0.79                                                                     & 0.69                                                                     & 0.62                                                                       & 0.74                                                             & 0.77                                                                       & 0.74             & 0.78          \\
\textbf{2DWNN}            & 0.93                 & 0.89               & \textbf{0.93}                 & 0.88               & 0.67             & \textbf{0.99}                                                                   & \textbf{0.99}                                                                     & \textbf{0.89}                                                                     & \textbf{0.94}                                                                       & \textbf{0.99}                                                                     & 0.62                                                                       & 0.82             & 0.88          \\
\textbf{STWNN}            & \textbf{0.99}          & \textbf{0.99} & \textbf{0.93} & \textbf{0.96} & 0.85 & 0.84               & 0.83                 & \textbf{0.89}                 & 0.92                   & 0.76                  & 0.75                   & \textbf{0.87} & \textbf{0.90}          \\ \bottomrule
\end{tabular}}
\vspace{-10pt}
\end{table*}


Table~\ref{tab:WVAR} also reports the skeleton-based classification based on the WiFi and video data on the WVAR dataset. Inspired by the work~\cite{zhao2018through,AAAAli2019making}, the skeletons derived from Alphapose~\cite{fang2017rmpe} are used to train the STWNN in LOS scenes. Based on the skeletons, the trained STWNN can further generate skeletons in non-line-of-sight (NLOS) scenes. Based on the skeletons, these results by SVM obtained an accuracy of 94$\%$, which is slightly lower than STWNN-based WiFi only. The reason behind this is mainly due to the distortions in predicted skeletons in NLOS scenes.

Table~\ref{tab:WAR} represents classification accuracy on the WAR dataset. STWNN outperforms all the others with over 93\% accuracy. STWNN achieves the best except for the action run. Regarding action run, STWNN has relatively 1\% lower accuracy than 2DWNN but still an absolute improvement of 16\% than LSTM~\cite{yousefi2017survey} in the original paper. Even in more complex actions of lie-down and fall, these improvements by STWNN compared with LSTM can also achieve $1\%$ and $5\%$, respectively.

These classification results on the HHI dataset are shown in Table~\ref{tab:HHI}. The OA of STWNN ranks the first ($90\%$). Especially, most of these results of 2DWNN and STWNN are better than all the other models, except for these results of hugging and punching-with-left-hand, but still are comparable with E2EDLF, which show that 2DWNN and STWNN are both very strong baselines.

These classification results on the CSLOS dataset are shown in Table~\ref{tab:HHI}. The classification accuracy of STWNN ranks first compared to all other two methods in two LOS scenes E1 and E2. As for E1, STWNN achieves the best results on all actions, with 2$\%$ higher than SVM~\cite{baha2021exploiting}. Concerning E2, the performance of STWNN is better except for no\_movement and walking which still are comparable with those of SVM~\cite{baha2021exploiting}. In other words, STWNN has good robustness in comparison to the other two models.

\begin{table}[h]
\centering
\vspace{-10pt}
\caption{Classification accuracy on the dataset CSLOS}
\label{tab:LOS}
\tiny
\begin{tabular}{l!{\vrule width \lightrulewidth}l!{\vrule width \lightrulewidth}ccccccc} 
\toprule
\textbf{Scenes}              & \textbf{Methods}  & \textbf{\begin{tabular}[c]{@{}c@{}}no\_ \\move\end{tabular}} & \textbf{falling} & \textbf{walking} & \textbf{\begin{tabular}[c]{@{}c@{}}sitting\_ \\standing\end{tabular}} & \textbf{turning} & \textbf{\begin{tabular}[c]{@{}c@{}}picking\_ \\up\end{tabular}}& \textbf{Average}  \\ 
\midrule
\multirow{4}{*}{\textbf{E1} }    & \textbf{SVM}~\cite{baha2021exploiting} & 0.98          & 0.86          & 1.00          & 0.91          & 0.90          & 0.92          & 0.94             \\
                & \textbf{2DWNN}    & 0.89          & 0.80          & 0.73          & 0.86          & 0.67          & 0.94          & 0.81             \\
                & \textbf{STWNN}    & \textbf{0.96} & \textbf{0.96} & \textbf{0.93} & \textbf{0.99} & \textbf{0.92} & \textbf{0.99} & \textbf{0.96}                 \\
                             &                   &             &             &             &             &             &             &                   \\
\multirow{4}{*}{\textbf{E2}}     & \textbf{SVM}~\cite{baha2021exploiting} & \textbf{0.95} & 0.82          & \textbf{0.99} & 0.82          & 0.81          & 0.82          & 0.89             \\
                & \textbf{2DWNN}    & 0.84          & 0.78          & 0.75          & 0.83          & 0.69          & 0.84          & 0.79             \\
                & \textbf{STWNN}    & 0.93          & \textbf{0.94} & 0.94          & \textbf{0.95} & \textbf{0.90} & \textbf{0.88} & \textbf{0.92}                     \\
\bottomrule
\end{tabular}
\vspace{-10pt}
\end{table}

\noindent\textbf{Qualitative Results} In this section, we show the effectiveness of CSI data on WVAR. Besides, we demonstrate the spatiotemporal scheme and the attention module of STWNN are meaningful at the feature level. 

Skeleton visualization is further to show the effectiveness of WVAR as mentioned in section 4.1. As seen in Fig.~\ref{fig:skeleton}(a)-(h), our STWNN yields robust skeletons close to Alphapose. Particularly, these partially covered actions such as kick are also well-estimated. This demonstrates that our CSI data on WVAR has a good efficiency in these two scenarios.

\begin{figure}[h!]
    \vspace{-5pt}
    \centering
    \includegraphics[scale=0.19]{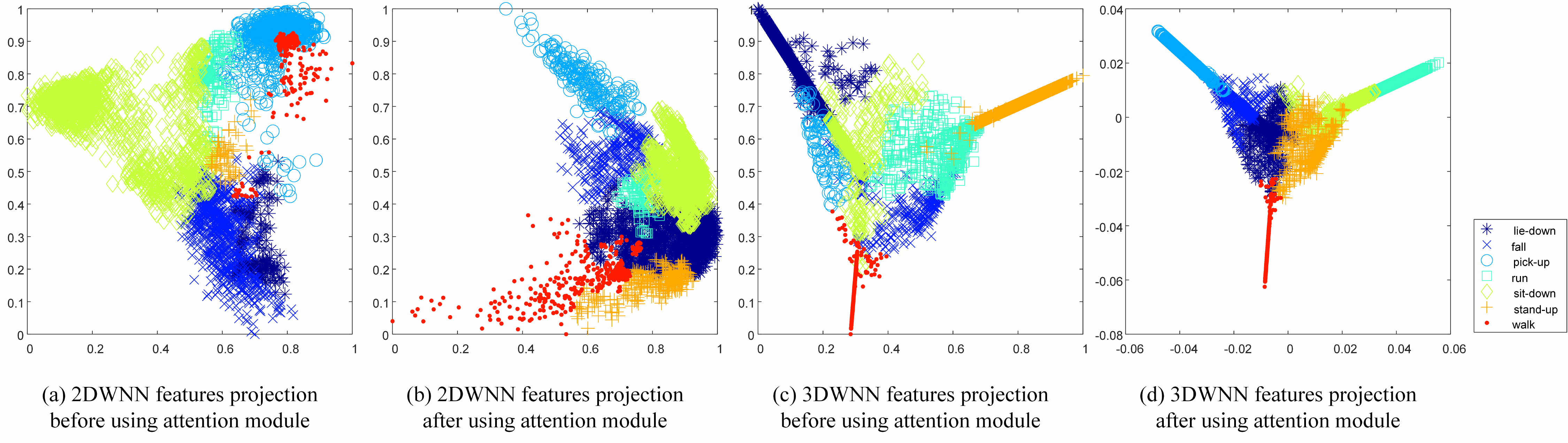}
    \caption{The features projections of the dataset WAR by 2DWNN and STWNN before and after using the attention module.}
    \label{fig:feature_attention}
\vspace{-15pt}
\end{figure}

In Fig.~\ref{fig:feature_attention}, we show gSOM~\cite{ilc2012generation} projections of the features of the WAR dataset by 2DWNN and 3DWNN before and after using the attention module, respectively. Features extracted from the same action tend to be near each other, and vice versa. Compared Fig.~\ref{fig:feature_attention}(a) with Fig.~\ref{fig:feature_attention}(c), we can find the features from the same action of 3DWNN are more compact than that of 2DWNN in terms of lie-down, walk, and stand-up. It proves that 3DWNN has better potential than 2DWNN in exploring the effective spatiotemporal features. Perceptually, comparing Fig.~\ref{fig:feature_attention}(c) with (d), features from the same category after using the attention module are more clustered than that before, such as run and pick-up. It further indicates that the attention module improves the efficiency of the features to a certain extent. 

\vspace{-10pt}
\section{Conclusion}
In this paper, an end-to-end spatiotemporal WiFi-based neural network STWNN was proposed to enhance the performance of privacy-preserving WiFi-based HAR. Its strength lay in the ability for the effective exploitation of the multi-scale spatiotemporal features and explicit maintenance of self-attention features. Moreover, we collected synchronous video and WiFi datasets WVAR to enable STWNN to see the skeleton in complex visual conditions like partial and full occlusions scenarios. In addition, we have compared the results of our proposed STWNN with the results of SVM, 2DWNN, and state-of-the-art competitors. The experiments on four benchmark datasets WVAR, WAR, HHI, and CSLOS showed that STWNN compares favorably against competitive baselines.





\bibliographystyle{IEEEbib}
\bibliography{3Drefs}

\begin{thebibliography}{10}

\bibitem{isogawa2020optical}
Mariko Isogawa, Ye~Yuan, Matthew O'Toole, and Kris~M Kitani,
\newblock ``Optical non-line-of-sight physics-based {3D} human pose
  estimation,''
\newblock in {\em CVPR}, 2020, pp. 7013--7022.

\bibitem{luo2021intelligent}
Yiyue Luo, Yunzhu Li, Michael Foshey, Wan Shou, Pratyusha Sharma, Tomas
  Palacios, Antonio Torralba, and Wojciech Matusik,
\newblock ``{Intelligent Carpet}: Inferring {3D} human pose from tactile
  signals,''
\newblock in {\em CVPR}, 2021, pp. 11255--11265.

\bibitem{wu2018wifi}
Bang Wu, Zixiang Ma, Stefan Poslad, and Yidong Li,
\newblock ``{WiFi} fingerprint based, indoor, location-driven activities of
  daily living recognition,''
\newblock in {\em BESC}. IEEE, 2018, pp. 148--151.

\bibitem{tan2018exploiting}
Bo~Tan, Qingchao Chen, Kevin Chetty, Karl Woodbridge, Wenda Li, and Robert
  Piechocki,
\newblock ``Exploiting {WiFi} channel state information for residential
  healthcare informatics,''
\newblock {\em IEEE Commun. Mag.}, vol. 56, no. 5, pp. 130--137, 2018.

\bibitem{wang2018csi}
Fei Wang, Jinsong Han, Shiyuan Zhang, Xu~He, and Dong Huang,
\newblock ``{CSI-net}: Unified human body characterization and action
  recognition,''
\newblock {\em arXiv:1810.03064}, 2018.

\bibitem{AAAAli2019making}
Tianhong Li, Lijie Fan, Mingmin Zhao, Yingcheng Liu, and Dina Katabi,
\newblock ``Making the invisible visible: {Action} recognition through walls
  and occlusions,''
\newblock in {\em ICCV}, 2019, pp. 872--881.

\bibitem{beddiar2020vision}
Djamila~Romaissa Beddiar, Brahim Nini, Mohammad Sabokrou, and Abdenour Hadid,
\newblock ``Vision-based human activity recognition: a survey,''
\newblock {\em Multimed. Tools. Appl.}, vol. 79, no. 41, pp. 30509--30555,
  2020.

\bibitem{li2021deep}
Chenning Li, Zhichao Cao, and Yunhao Liu,
\newblock ``Deep {AI} enabled ubiquitous wireless sensing: A survey,''
\newblock {\em CSUR}, vol. 54, no. 2, pp. 1--35, 2021.

\bibitem{fang2020person}
Shiwei Fang, Sirajum Munir, and Shahriar Nirjon,
\newblock ``Person tracking and identification using cameras and {Wi-Fi}
  channel state information {(CSI)} from smartphones: dataset,''
\newblock in {\em SenSys}, 2020, pp. 26--30.

\bibitem{alazrai2020dataset}
Rami Alazrai, Ali Awad, Alsaify Baha’A, Mohammad Hababeh, and Mohammad~I
  Daoud,
\newblock ``A dataset for {Wi-Fi-based} human-to-human interaction
  recognition,''
\newblock {\em Data Brief}, vol. 31, pp. 105668, 2020.

\bibitem{wang2016rt}
Hao Wang, Daqing Zhang, Yasha Wang, Junyi Ma, Yuxiang Wang, and Shengjie Li,
\newblock ``{RT-Fall}: A real-time and contactless fall detection system with
  commodity {WiFi} devices,''
\newblock {\em IEEE Trans. Mobile Comput.}, vol. 16, no. 2, pp. 511--526, 2016.

\bibitem{baha2020dataset}
Alsaify Baha’A, Mahmoud~M Almazari, Rami Alazrai, and Mohammad~I Daoud,
\newblock ``A dataset for {Wi-Fi}-based human activity recognition in
  line-of-sight and non-line-of-sight indoor environments,''
\newblock {\em Data in Brief}, vol. 33, pp. 106534, 2020.

\bibitem{yousefi2017survey}
Siamak Yousefi, Hirokazu Narui, Sankalp Dayal, Stefano Ermon, and Shahrokh
  Valaee,
\newblock ``A survey on behavior recognition using {WiFi} channel state
  information,''
\newblock {\em IEEE Commun. Mag.}, vol. 55, no. 10, pp. 98--104, 2017.

\bibitem{crammer2001algorithmic}
Koby Crammer and Yoram Singer,
\newblock ``On the algorithmic implementation of multiclass kernel-based vector
  machines,''
\newblock {\em J. Mach. Learn Res.}, vol. 2, no. Dec, pp. 265--292, 2001.

\bibitem{ho1995random}
Tin~Kam Ho,
\newblock ``Random decision forests,''
\newblock in {\em ICDAR}. IEEE, 1995, vol.~1, pp. 278--282.

\bibitem{eddy2004hidden}
Sean~R Eddy,
\newblock ``What is a hidden {Markov} model?,''
\newblock {\em Nat. Biotechnol.}, vol. 22, no. 10, pp. 1315--1316, 2004.

\bibitem{szegedy2015going}
Christian Szegedy, Wei Liu, Yangqing Jia, Pierre Sermanet, Scott Reed, Dragomir
  Anguelov, Dumitru Erhan, Vincent Vanhoucke, and Andrew Rabinovich,
\newblock ``Going deeper with convolutions,''
\newblock in {\em CVPR}, 2015, pp. 1--9.

\bibitem{he2016deep}
Kaiming He, Xiangyu Zhang, Shaoqing Ren, and Jian Sun,
\newblock ``Deep residual learning for image recognition,''
\newblock in {\em CVPR}, 2016, pp. 770--778.

\bibitem{iandola2016squeezenet}
Forrest~N Iandola, Song Han, Matthew~W Moskewicz, Khalid Ashraf, William~J
  Dally, and Kurt Keutzer,
\newblock ``{SqueezeNet}: {AlexNet}-level accuracy with 50x fewer parameters
  and< 0.5 mb model size,''
\newblock {\em arXiv:1602.07360}, 2016.

\bibitem{alazrai2020end}
Rami Alazrai, Mohammad Hababeh, Alsaify Baha’A, Mostafa~Z Ali, and Mohammad~I
  Daoud,
\newblock ``An end-to-end deep learning framework for recognizing
  human-to-human interactions using {Wi-Fi} signals,''
\newblock {\em IEEE Access}, vol. 8, pp. 197695--197710, 2020.

\bibitem{zhao2018through}
Mingmin Zhao, Tianhong Li, Mohammad Abu~Alsheikh, Yonglong Tian, Hang Zhao,
  Antonio Torralba, and Dina Katabi,
\newblock ``Through-wall human pose estimation using radio signals,''
\newblock in {\em CVPR}, 2018.

\bibitem{fang2017rmpe}
Hao-Shu Fang, Shuqin Xie, Yu-Wing Tai, and Cewu Lu,
\newblock ``{RMPE}: Regional multi-person pose estimation,''
\newblock in {\em ICCV}, 2017, pp. 2334--2343.

\bibitem{baha2021exploiting}
Alsaify Baha’A, Mahmoud~M Almazari, Rami Alazrai, and Mohammad~I Daoud,
\newblock ``Exploiting {Wi-Fi} signals for human activity recognition,''
\newblock in {\em ICICS}. IEEE, 2021, pp. 245--250.

\bibitem{ilc2012generation}
Nejc Ilc and Andrej Dobnikar,
\newblock ``Generation of a clustering ensemble based on a gravitational
  self-organising map,''
\newblock {\em Neurocomputing}, vol. 96, pp. 47--56, 2012.

\end{thebibliography}

\end{document}